\documentclass[12pt]{article}
\usepackage{amscd}
\usepackage{amsfonts}
\usepackage{CJK}
\usepackage{amssymb}
\usepackage{amsmath}
\usepackage{color}
\usepackage{cite}
\usepackage{bbm}
\usepackage{graphicx}
\usepackage{cite}
\usepackage{setspace}

\newtheorem{theorem}{Theorem}[section]
\newtheorem{prop}[theorem]{Proposition}
\newtheorem{cor}[theorem]{Corollary}

\newtheorem{define}[theorem]{Definition}
\newtheorem{remark}[theorem]{Remark}

\newcommand\btd{\raise 2pt \hbox{$\hat\bigtriangledown$}\hskip 1.5pt}
\newcommand\bt{\raise 2pt \hbox{$\bigtriangledown$}\hskip 1.5pt}

\def\no{\nonumber}

\begin{document}
\title{Determination  of approximate nonlinear self-adjointenss and
application to construct approximate conservation law}
\author{Zhi-Yong Zhang\footnote{E-mail: zhiyong-2008@163.com; Tel:+86 010 88803103 } %\ \ \ \  \ \ \ \ \ \ \ \ \ \ \ \
\\\small~College of Sciences, North China University of
Technology, Beijing 100144, P.R. China}
\date{}
\maketitle

\noindent{\bf Abstract:} Approximate nonlinear self-adjointness is
an effective method to construct approximate conservation law of
perturbed partial differential equations (PDEs). In this paper, we
study the relations between approximate nonlinear self-adjointness
of perturbed PDEs and nonlinear self-adjointness of the
corresponding unperturbed PDEs, and consequently provide a simple
approach to discriminate approximate nonlinear self-adjointness of
perturbed PDEs. Moreover, a succinct approximate conservation law
formula by virtue of the known conservation law of the unperturbed
PDEs is given in an explicit form. As an application, we classify a
class of perturbed wave equations to be approximate nonlinear
self-adjointness and construct the general approximate conservation
laws formulae. The specific examples demonstrate that approximate
nonlinear self-adjointness can generate new approximate conservation
laws.

 %\noindent{\bf PACS numbers:} 02.20.Sv, 02.30.Jr, 02.30.Mv


\section{Introduction}
The discovery of conservation laws played a pivotal role in the
comprehensive study of the properties and solutions of PDEs arising
in the applied sciences such as physical chemistry, fluid mechanics,
etc. For example, the knowledge of conservation laws is useful in
the numerical integration of PDEs such as controlling numerical
errors. An infinite sequence of conserved densities is a predictor
of the existence of solitons  and complete integrability which means
that the PDEs can be solved with the inverse scattering transform
method \cite{ab-1991,ma-1988}. Hence, in order to achieve these
goals, one needs efficient methods to compute conservation laws.

Different types of PDEs have their distinct structures and thus
should be studied by different techniques. By considering whether
the PDEs involve a perturbed parameter or not, PDEs are divided into
two types which are unperturbed PDEs and perturbed one, thus
researchers adopted different methods to construct conservation laws
of the two types of PDEs
\cite{blu1,Olver,ba-1996}. %Among various approaches, finding conservation laws
%plays an important role  such as detecting the integrability,
%finding solutions, linearization, connecting with modern numerical
%methods, exploring properties of the PDEs (formulation of familiar
%physical laws such as for mass, energy and momentum) and so on
%.

For unperturbed PDEs, Noether theorem proposed an elegant and
constructive way to find conservation laws with the admitted
symmetry if the PDEs can be written in Euler-Lagrange form
\cite{noether}. The prerequisite of applying Noether theorem is that
the PDEs possess a variational structure, that means the method
losses its effectiveness when dealing with the PDEs without having a
Lagrangian, thus the methods which are less restriction on the
existence of Lagrangian come into being. For example, in
\cite{kara-2006}, Kara and Mahomed presented a partial Noether
approach for Euler-Lagrange type equations.  Anco and Bluman
proposed multiplier method to construct local conservation laws for
PDEs expressed in a standard Cauchy-Kovalevskaya form, which did not
count on the existence of Lagrangian \cite{blu11,blu21}.  Recently,
Ibragimov suggested the general concept of nonlinear
self-adjointness whose idea can be traced back to
\cite{ba-1931,ca-1988}, which embraces strict self-adjointness, qusi
self-adjointness and weak self-adjointness presented earlier
\cite{ib-2011,nhi,ib-2007,nh-2006,nh-2007,gan-2011}. Nonlinear
self-adjointness and its subclasses overcame the restriction of
Lagrangian and have exerted effectiveness when constructing
conservation laws of some PDEs
\cite{Fs-2012,fre-2013,tt-2012,gan-2012,nhi-2011}.

The low efficiency of classical symmetry method for some perturbed
PDEs, where the involved perturbed parameter disturbed the symmetry
properties of the unperturbed PDEs and even made the PDEs possess
few symmetries, promoted the emergence of approximate symmetry.
Consequently, two well-known approaches originated from Baikov et
al. \cite{ba-1991} and Fushchich and Shtelen \cite{fs-1989} arose,
which employed perturbation techniques on the symmetry operator and
dependent variables to obtain approximate symmetry respectively.
Comparisons about the two methods were performed by considering
several perturbed PDEs in \cite{pak,ron}. The extended applications
and forms of approximate symmetry such as approximate conditional
symmetry and approximate potential symmetry can be found in
\cite{qu1,qu2,eu-1994} and references therein.

In order to construct approximate conservation laws with approximate
symmetry, it is natural to extend the procedures for unperturbed
PDEs to tackle perturbed PDEs. In \cite{ah-1999,ah}, the authors
studied how to construct approximate conservation laws for perturbed
PDEs with approximate Lie-B\"{a}cklund symmetries. In
\cite{ag-2006}, a basis of approximate conservation laws for
perturbed PDEs was discussed. Johnpillai et.al \cite{ah-2009} showed
how to find approximate conservation laws via approximate Noether
type symmetry operators associated with partial Lagrangian.

Quite recently, we introduced the notion of approximate nonlinear
self-adjointness to construct approximate conservation laws of
perturbed PDEs, which extended the results in unperturbed case and
the ones of perturbed ordinary differential equations
\cite{zhang-2013}. As an equivalent form of the definition, we gave
a computable formula to discriminate approximate nonlinear
self-adjointness and applied it to find approximate conservation
laws of a class of perturbed nonlinear wave equation
\cite{zhang-2013}. However, some perturbed PDEs are not
approximately nonlinearly self-adjoint but we know the final results
only by using the proposed formula to check it and finally obtain
trivial substitution, which expends large time and computations.

The purpose of this paper aims at two aspects:

1. Study the relations between approximate nonlinear
self-adjointness of perturbed PDEs and nonlinear self-adjointness of
the corresponding unperturbed PDEs, and consequently provide a
simple approach to discriminate approximate nonlinear
self-adjointness of perturbed PDEs.

2. We also give an explicit approximate conservation law formula by
virtue of the known conservation law of the corresponding
unperturbed PDEs.
%Applications to several perturbed PDEs illustrate the effectiveness
%of the results.

The rest of the paper is arranged as follows. In Section 2, some
related basic notions and principles are reviewed. Section 3
concentrates on the main results of the paper. In Section 4, we
consider a class of perturbed wave equations and construct the
corresponding approximate conservation law formulae. The last
section contains conclusion and discussion of our work.

\section{Preliminaries}
In this section, we recall some notions and principles related with
nonlinear self-adjointness and approximate nonlinear
self-adjointness.
\subsection{Notations}
We first define some notations which will be used throughout the
paper.

Let $x=(x^1,\dots,x^n)$ be an independent variable set,
$u=(u^1,\dots,u^m)$, $v=(v^1,\dots,v^m)$ and $w=(w^1,\dots,w^m)$ as
three dependent variable sets, $\phi(x,u)=(\phi^{1},\dots,\phi^m)$,
and $\psi(x,u)=(\psi^{1},\dots,\psi^m)$
$\varphi(x,u)=(\varphi^{1},\dots,\varphi^m)$ as three
$m$-dimensional vector functions, whose different order derivatives
are denoted as follows
\begin{eqnarray}
&&\no
\Lambda_{(1)}=\{\Lambda^{\sigma}_{i_1}\},~~\Lambda_{(2)}=\{\Lambda^{\sigma}_{i_1i_2}\},~~\dots,
~~\Lambda_{(r)}=\{\Lambda^{\sigma}_{i_1\dots i_r}\},
\end{eqnarray}
where $\Lambda^{\sigma}_{i_1\dots i_s}=D_{i_1}D_{i_2}\dots
D_{i_s}(\Lambda^{\sigma})$ and $\sigma=1,2,\dots,m$. The symbol
``$\Lambda$" denotes the dependent variables $u,v,w$ and vector
functions $\phi(x,u),\psi(x,u),\varphi(x,u)$. Hereinafter, $D_i$
denotes the total derivative operator with respect to $x^i$. For
example, for one dependent variable $u=u(x,t)$ with $x=x^1,t=x^2$,
one has
\begin{eqnarray}
&&\no D_x = \frac{\partial}{\partial x} +u_x
\frac{\partial}{\partial u} +u_{xx} \frac{\partial}{\partial u_x}
+u_{xt}\frac{\partial}{\partial u_t} +\dots.
\end{eqnarray}
Note that the summation convention for repeated indices is used
throughout the paper if no special notations are added.

Next, we take the following system of $m$ PDEs with $r$th-order
\begin{eqnarray}\label{perturbnew}
E_{\alpha} = E_{\alpha}^0(x,u,u_{(1)},\cdots,u_{(r)})+ \epsilon
E_{\alpha} ^1(x,u,u_{(1)},\cdots,u_{(r)})=0,
\end{eqnarray}
to recall some related notions and principles, where $\epsilon$ is a
small parameter and $\alpha=1,2,\dots,m$. System (\ref{perturbnew})
is called perturbed PDEs while the system which does not contain the
perturbed term $\epsilon E_{\alpha} ^1(x,u,u_{(1)},\cdots,u_{(r)})$
in the form
\begin{eqnarray}\label{perturb}
E_{\alpha}^0(x,u,u_{(1)},\cdots,u_{(r)})=0,
\end{eqnarray}
is named by unperturbed PDEs.

\subsection{Approximate symmetry}
In this subsection, we briefly introduce two methods to find
approximate symmetry of perturbed PDEs, which will be used to
construct approximate conservation laws.

The first method is proposed by Fushchich and Shtelen, which employs
a perturbation of dependent variables, that is, expanding the
dependent variable with respect to the small parameter $\epsilon$
yields
\begin{eqnarray}\label{tran}
&& u^\sigma =\sum^{\infty}_{k=0}\epsilon^k u^{\sigma_k},\qquad
0<\epsilon \ll 1,
\end{eqnarray}
where $u^{\sigma_k}$ are new introduced dependent variables, after
inserting expansion (\ref{tran}) into system (\ref{perturbnew}),
then approximate symmetry is defined as the exact symmetry of the
system corresponding to each order in the small parameter
$\epsilon$. We refer to reference \cite{fs-1989} for further details
about the method.

The second approach, initiated by Baikov et al., is no perturbation
of the dependant variables but a perturbation of the symmetry
generator \cite{ba-1991}.

A first-order approximate symmetry of system (\ref{perturbnew}),
with the infinitesimal operator form $X = X_0 +\epsilon X_1$,  is
obtained by solving for $X_1$ in
\begin{eqnarray}\label{second}
&& X_1(E_{\alpha}^0)_{\mid_{E_{\alpha}^0=0}}+H=0,
\end{eqnarray}
where the auxiliary function $H$ is obtained by
\begin{eqnarray}\label{auxiliary}
\no H=\frac{1}{\epsilon}X_0(E_{\alpha})_{\mid_{E_{\alpha}=0}}.
\end{eqnarray}
$X_0$ is an exact symmetry of unperturbed PDEs $E_{\alpha}^0=0$. The
notation $|_{\Delta=0}$, hereinafter, means evaluation on the
solution manifold of $\Delta=0$.

Note that  in this paper, the approximate symmetries used to
construct approximate conservation laws are computed by the method
originated from Baikov et.al \cite{ba-1991}.
\subsection{Nonlinear self-adjointness}
The main idea of nonlinear self-adjointness is to turn the system of
PDEs into Lagrangian equations by artificially adding additional
variables, then apply the conservation law theorem proved in
\cite{nh-2007} to construct conservation laws.

 Let $\mathcal {L}$ be the formal
Lagrangian of system (\ref{perturb}) given by
\begin{eqnarray}\label{lagrangian}
&& \mathcal {L} = v^{\beta}E_{\beta}^0(x,u,u_{(1)},\cdots,u_{(r)}),
\end{eqnarray}
then the adjoint equations of system (\ref{perturb}) are defined by
\begin{eqnarray}\label{adequation}
(E_{\alpha}^0)^{\ast}(x,u,v,u_{(1)},v_{(1)},\cdots,u_{(r)},v_{(r)})=\frac{\delta\mathcal
{L}}{\delta u^{\sigma}}=0,
\end{eqnarray}
where $\delta/\delta u^{\sigma}$ is the Euler-Lagrange operator
written as
 \begin{eqnarray}\label{var-op}
\frac{\delta}{\delta u^{\sigma}}=\frac{\partial}{\partial
u^{\sigma}}+\sum_{s=1}^{\infty}(-1)^sD_{i_1}\dots
D_{i_s}\frac{\partial}{\partial u^{\sigma}_{i_1\dots i_s}}.
\end{eqnarray}
\begin{define} \label{def-non}(Nonlinear self-adjointness \cite{ib-2011}) The
system (\ref{perturb}) is said to be nonlinearly self-adjoint if the
adjoint system (\ref{adequation}) is satisfied for all solutions $u$
of system (\ref{perturb}) upon a substitution $v=\varphi(x,u)$ such
that $\varphi(x,u)\neq 0$.\end{define}

Here, $\varphi(x,u)\neq 0$  means that not all elements of
$\varphi(x,u)$ equal zero. Definition \ref{def-non} is equivalent to
the following identity holding for the undetermined parameters (or
functions) $\lambda_{\alpha}^{\beta}$
\begin{eqnarray}\label{equ-id1}
&&
(E_{\alpha}^0)^{\ast}(x,u,v,u_{(1)},v_{(1)},\cdots,u_{(r)},v_{(r)})_{|_{v=\varphi}}
=\lambda_{\alpha}^{\beta}E_{\beta}^0(x,u,u_{(1)},\cdots,u_{(r)}),
\end{eqnarray}
or $[\delta (v^\beta E^0_\beta)/\delta
u^{\sigma}]_{|_{v=\varphi}}=\lambda_{\alpha}^{\beta}E_{\beta}^0$,
which is applicable in the computations.
\subsection{Approximate nonlinear self-adjointness}
In \cite{zhang-2013}, we proposed the definition of approximate
nonlinear self-adjointness and extended the above procedure to deal
with perturbed PDEs.
%Hereinafter, $\varphi_{(i)}$ (resp. $\phi_{(i)}$ below), similar as
%$v_{(i)}$ and $u_{(i)}$, stands for all $i$th-order partial
%derivatives of $\varphi$ (resp. $\phi$) with respect to $x$.
The formal Lagrangian $\mathcal {\widetilde{L}}$ of perturbed system
(\ref{perturbnew}) is given by
\begin{eqnarray}\label{lagrangian1}
\no \mathcal {\widetilde{L}} =
w^{\beta}[E_{\beta}^0(x,u,u_{(1)},\cdots,u_{(r)})+ \epsilon\,
E_{\beta}^1(x,u,u_{(1)},\cdots,u_{(r)})],
\end{eqnarray}
then the adjoint equations of system (\ref{perturbnew}) are written
as
\begin{eqnarray}\label{adequationper}
E_{\alpha}^{\ast}(x,u,w,u_{(1)},w_{(1)},\dots,u_{(r)},w_{(r)})=\frac{\delta\mathcal
{\widetilde{L}}}{\delta u^{\sigma}}=0.
\end{eqnarray}

\begin{define} \label{def-1}(Approximate nonlinear self-adjointness \cite{zhang-2013}) The
perturbed system (\ref{perturbnew}) is called approximate nonlinear
self-adjointness if the adjoint system (\ref{adequationper}) is
approximate satisfied for all solutions $u$ of system
(\ref{perturbnew}) upon a substitution
$w=\psi(x,u)+\epsilon\phi(x,u),$ such that not both $\psi$ and
$\phi$ are identically equal to zero.\end{define}

Definition \ref{def-1} extends the results of unperturbed case and
perturbed ordinary differential equations in \cite{kara-2002} and is
equivalent to the following identity with undetermined parameters
(or functions) $\lambda_\alpha^\beta$ and $\mu_\alpha^\beta$,
\begin{eqnarray}\label{forper}
&& \no
\hspace{-0.5cm}E_{\alpha}^{\ast}(x,u,w,u_{(1)},w_{(1)},\dots,u_{(r)},w_{(r)})_{|_{w=\psi+\epsilon
\phi}}\\
&&
\hspace{0.5cm}-\left[(\lambda_\alpha^\beta+\epsilon\mu_\alpha^\beta)
E^0_{\beta}(x,u,u_{(1)},\dots,u_{(r)})+\epsilon \lambda_\alpha^\beta
E^1_{\beta}(x,u,u_{(1)},\dots,u_{(r)})\right]=O(\epsilon^2),
\end{eqnarray}
which provides a computable equality to discriminate approximate
nonlinear self-adjointness of perturbed PDEs \cite{zhang-2013}.

The following properties about approximate nonlinear
self-adjointness of perturbed PDEs will be used in the proof of the
main results in Section \ref{sub-1}.
\begin{theorem} \label{th-2}  \cite{zhang-2013} If adjoint system (\ref{adequation}) has
solutions in the form $v^{\sigma}=\epsilon f^{\sigma}(x,u)$ with
some functions $f^{\sigma}(x,u)$, then system (\ref{perturbnew}) is
approximately nonlinearly self-adjoint.\end{theorem}
\subsection{Approximate conservation law}
$k$th-order approximate conservation law is defined as follows.

\begin{define}(Approximate conservation law \cite{ba-1996})
The vector $T = (T^1 , T^2 ,\dots T^n)$ defined by $T^i =
T^i_0+\epsilon T^i_1+\dots+\epsilon^kT^i_k$ is approximate conserved
vector of system (\ref{perturbnew}) if $T^i$ satisfies approximate
equation $D_i T^i = O(\epsilon^{k+1})$ for all solutions of system
(\ref{perturbnew}), which defines a $k$th-order approximate
conservation law.\end{define}

The major consideration of the paper concentrates on the first-order
approximate conservation law which is defined as $D_i T^i =
O(\epsilon^2)$ with $T^i=T^i_0+\epsilon T^i_1$.

The following theorem will be used to construct conservation laws
for both unperturbed and perturbed cases \cite{nh-2007}.

\begin{theorem} \label{con-law} Any infinitesimal symmetry (Local and nonlocal)
\begin{eqnarray}
&&\no X=\xi^i(x,u,u_{(1)},\dots)\frac{\partial}{\partial
x^i}+\eta^{\sigma}(x,u,u_{(1)},\dots)\frac{\partial}{\partial
u^{\sigma}}
\end{eqnarray}
of system (\ref{perturb}) leads to a conservation law $D_i(C^i)=0$
constructed by the formula
\begin{eqnarray}\label{formula}
&&\no C^i=\xi^i\mathcal {K}+W^{\sigma}\Big[\frac{\partial \mathcal
{K}}{\partial u_i^{\sigma}}-D_j(\frac{\partial \mathcal
{K}}{\partial u_{ij}^{\sigma}})+D_jD_k(\frac{\partial \mathcal
{K}}{\partial u_{ijk}^{\sigma}})-\dots\Big]\\&&\hspace{1cm}
+D_j(W^{\sigma}) \Big[\frac{\partial \mathcal {K}}{\partial
u_{ij}^{\sigma}}-D_k(\frac{\partial \mathcal {K}}{\partial
u_{ijk}^{\sigma}})+\dots\Big]+D_jD_k(W^{\sigma})\Big[\frac{\partial
\mathcal {K}}{\partial u_{ijk}^{\sigma}}-\dots\Big]+\dots,
\end{eqnarray}
where $W^{\sigma}=\eta^{\sigma}-\xi^ju_j^{\sigma}$ and $\mathcal
{K}$ is the formal Lagrangian. \end{theorem}

Generally speaking, the term $\xi^i\mathcal {K}$ with $\mathcal {K}$
in the form (\ref{lagrangian}) can be omitted because it vanishes
identically on the solution manifold of the studying PDEs.

\section{Main results}\label{sub-1}
In this section, we first study the relations between approximate
nonlinear self-adjointness of perturbed system (\ref{perturbnew})
and nonlinear self-adjointness of unperturbed PDEs (\ref{perturb}),
then give a discriminant criterion of approximate nonlinear
self-adjointness and a succinct approximate conservation law formula
for perturbed system (\ref{perturbnew}).

\begin{theorem} \label{th-5} If unperturbed system (\ref{perturb}) is
nonlinearly self-adjoint, then perturbed system (\ref{perturbnew})
is approximately nonlinearly self-adjoint.
 \end{theorem}

\emph{Proof.} If system (\ref{perturb}) is nonlinearly self-adjoint,
then by Definition \ref{def-non}, for all solutions $u$ of system
(\ref{perturb}), there exists a substitution $v=\varphi(x,u)\neq 0$
which solves adjoint equations (\ref{adequation}). Since adjoint
system (\ref{adequation}) is linear in $v$ and its derivatives, thus
$\epsilon v=\epsilon\varphi(x,u)$ also satisfies system
(\ref{adequation}). Then by Theorem \ref{th-2}, system
(\ref{perturbnew}) is approximately nonlinearly self-adjoint. The
proof ends. $\hfill{} \blacksquare$

Conversely, we have the following results.
\begin{theorem} \label{th-3} If  perturbed system (\ref{perturbnew}) is approximately
nonlinearly self-adjoint, and equations $\lambda_\alpha^\beta
E^0_{\beta}=0$ for $\lambda_\alpha^\beta$ only have zero solutions
$\lambda_\alpha^\beta=0\,(\alpha,\beta=1,\dots,m)$, then unperturbed
system (\ref{perturb}) is nonlinearly self-adjoint. \end{theorem}

\emph{Proof.} If perturbed system (\ref{perturbnew}) is
approximately nonlinearly self-adjoint, then by Definition
\ref{def-1}, there exists a nontrivial substitution
$w=\psi(x,u)+\epsilon\phi(x,u)$ such that equality (\ref{forper})
holds.

Observe that $\mathcal {\widetilde{L}}=w^{\beta}E_{\beta}^0+\epsilon
w^{\beta}E_{\beta}^1$, then system (\ref{adequationper}) becomes
\begin{eqnarray}\label{proof-1}
&&\frac{\delta\mathcal {\widetilde{L}}}{\delta
u^{\sigma}}=\frac{\delta(w^{\beta}E_{\beta}^0)}{\delta
u^{\sigma}}+\epsilon \frac{\delta (w^{\beta}E_{\beta}^1)}{\delta
u^\sigma}.
\end{eqnarray}

Inserting (\ref{proof-1}) and $w=w_0+\epsilon w_1$ into
(\ref{forper}), one has
\begin{eqnarray}\label{forper-proof}
&& \no \hspace{-0.5cm}\frac{\delta (w_0^{\beta}E_{\beta}^0)}{\delta
u^{\sigma}}_{|_{w_0=\psi}}+\epsilon \left[\frac{\delta
(w_0^{\beta}E_{\beta}^1)}{\delta
u^\sigma}_{|_{w_0=\psi}}+\frac{\delta
(w_1^{\beta}E_{\beta}^0)}{\delta
u^\sigma}_{|_{w_1=\phi}}\right]-\left[(\lambda_\alpha^\beta+\epsilon\mu_\alpha^\beta)
E^0_{\beta}+\epsilon \lambda_\alpha^\beta
E^1_{\beta}\right]=O(\epsilon^2),
\end{eqnarray}
and then separating it with respect to the perturbed parameter
$\epsilon$ up to first order, we obtain
\begin{eqnarray}\label{proof-2}
&&\no \epsilon^0:~~\frac{\delta (w_0^{\beta}E_{\beta}^0)}{\delta
u^{\sigma}}_{|_{w_0=\psi}}=\lambda_\alpha^\beta
E^0_{\beta},\\
&&\epsilon^1:~~\frac{\delta (w_0^{\beta}E_{\beta}^1)}{\delta
u^\sigma}_{|_{w_0=\psi}}+\frac{\delta
(w_1^{\beta}E_{\beta}^0)}{\delta
u^\sigma}_{|_{w_1=\phi}}=\mu_\alpha^\beta
E^0_{\beta}+\lambda_\alpha^\beta E^1_{\beta}.
\end{eqnarray}
Obviously, the first equation in (\ref{proof-2}) is just the
required condition (\ref{equ-id1}) for nonlinear self-adjointness of
the unperturbed PDEs (\ref{perturb}).

Now we assume that system (\ref{perturb}) is not nonlinearly
self-adjoint, that means the first equation in (\ref{proof-2}) has
no nontrivial solutions, i.e., $\psi(x,u)=0$. Since equations
$\lambda_\alpha^\beta E^0_{\beta}=0$ for $\lambda_\alpha^\beta$ only
have zero solutions
$\lambda_\alpha^\beta=0\,(\alpha,\beta=1,\dots,m)$ and $\delta
(w_0^{\beta}E_{\beta}^1)/\delta u^\sigma$ is linear in $w_0$ and its
derivatives, thus $[\delta (w_0^{\beta}E_{\beta}^1)/\delta
u^\sigma]_{|_{w_0=\psi=0}}=0$ and the second equation becomes
\begin{eqnarray}\label{proof-3}
 &&\frac{\delta
(w_1^{\beta}E_{\beta}^0)}{\delta
u^\sigma}_{|_{w_1=\phi}}=\mu_\alpha^\beta E^0_{\beta}.
\end{eqnarray}

Except for the introduced variable $w_1$ and unknown function
$\mu_\alpha^\beta$, system (\ref{proof-3}) has the same form as the
first equation in (\ref{proof-2})  and thus only has zero solution
$\phi(x,u)=0$, then the required substitution
$w=\psi(x,u)+\epsilon\phi(x,u)=0$, that means system
(\ref{perturbnew}) is not approximately nonlinearly self-adjoint,
which is a contradiction. Hence, unperturbed PDEs (\ref{perturb}) is
nonlinearly self-adjoint. It proves the theorem. $\hfill{}
\blacksquare$
%
%For instance, equation $F=u_{tt}-u_{xx}+\epsilon uu_t=0$ is showed
%to be approximate nonlinear self-adjointness \cite{zhang-2013}.
%Here, the adjoint equation of linear unperturbed equation
%$u_{tt}-u_{xx}=0$ is nonlinearly self-adjoint, thus by means of
%Theorem \ref{th-3}, equation $F=0$ is approximately nonlinearly
%self-adjoint.

Theorem \ref{th-5} and Theorem \ref{th-3} build a connection between
approximate nonlinear self-adjointness of perturbed PDEs and
nonlinear self-adjointness of its corresponding unperturbed PDEs and
thus provides a shortcut to discriminate approximate nonlinear
self-adjointness of perturbed PDEs, which attributes the
determination of approximate nonlinear self-adjointness of perturbed
PDEs to discriminate nonlinear self-adjointness of its corresponding
unperturbed system. %Meanwhile, Theorem  \ref{th-3} only presents a
%determination criterion and does not give the specific
\begin{remark}
The condition that system $\lambda_\alpha^\beta E^0_{\beta}=0$ for
$\lambda_\alpha^\beta$ only has zero solutions
$\lambda_\alpha^\beta=0\,(\alpha,\beta=1,\dots,m)$ holds if each
$E^0_{\beta}=0$ has a distinct term which does not appear in other
equations. Meanwhile, nonzero solutions for $\lambda_\alpha^\beta
E^0_{\beta}=0$ may arise when  two or more equations $E^0_{\beta}=0$
are identical.
\end{remark}

For example, equation $\lambda_\alpha^1 E^0_{1}+\lambda_\alpha^2
E^0_{2}+\lambda_\alpha^3 E^0_{3}=0$, where $\alpha=1,2,3$ and
$E^0_{1}=u_t+vu_x+uw_x=0,E^0_{2}=v_t+vu_x+uw_x=0$ and
$E^0_{3}=w_t+vu_x+uw_x=0$, then $u_t$ in $ E^0_{1}=0$, $v_{t}$ in $
E^0_{2}=0$ and $w_t$ in $ E^0_{3}=0$ are distinct and do not appear
in other equations, thus
$\lambda_\alpha^1=\lambda_\alpha^2=\lambda_\alpha^3=0$. On the other
hand, if $E^0_{1}=E^0_{2}=E^0_{3}$, then $(\lambda_\alpha^1
+\lambda_\alpha^2 +\lambda_\alpha^3) E^0_{1}=0$, thus
$\lambda_\alpha^1 =-\lambda_\alpha^2 -\lambda_\alpha^3$ which may
provide nonzero solutions for $\lambda_\alpha^\beta(\beta=1,2,3)$.

In particular, for a scalar perturbed PDE, we have the following
sufficient and necessary condition for the relations.
\begin{theorem} \label{th-4} A scalar perturbed PDE
is approximately nonlinearly self-adjoint if and only if its
corresponding unperturbed PDE is nonlinearly
self-adjoint.\end{theorem}

\emph{Proof.} The sufficiency is a direct result of Theorem
\ref{th-5} for $m=1$, thus we only show necessity.

Let the scalar perturbed PDE be in the form $E=E^0+\epsilon\,E^1$.
In this case, the first equation in (\ref{proof-2}) becomes
\begin{eqnarray}
\no \frac{\delta(w_0E^0)}{\delta u}_{|_{w_0=\psi}}=\lambda \, E^0.
\end{eqnarray}

Now assume that unperturbed PDE is not nonlinearly self-adjoint,
i.e., $w_0=\psi=0$, then equation $\lambda \, E^0=0$ only has
solution $\lambda=0$, then following the idea of the proof of
Theorem \ref{th-3}, one obtains that the substitution for
approximate nonlinear self-adjointness is
$w=\psi(x,u)+\epsilon\phi(x,u)=0$, which contradicts with the
approximate nonlinear self-adjointness of $E=0$, so unperturbed PDE
is nonlinearly self-adjoint.$\hfill{} \blacksquare$

For the perturbed PDEs sharing the same unperturbed PDE, we have the
following results.

\begin{cor}\label{cor-2}
If a scalar PDE is not nonlinearly self-adjoint, then the perturbed
system consisting of the PDE with any perturbed terms is not
approximately nonlinearly self-adjoint.
\end{cor}

For example, we consider the short pulse equation
\begin{equation}\label{short}
u_{xt}=u+\frac{1}{6}(u^3)_{xx},
\end{equation}
which was suggested in \cite{sch-2004} as a mathematical model for
the propagation of ultra-short light pulses in media with
nonlinearities. Eq.(\ref{short}) is not nonlinearly self-adjoint
with the substitution in the form $\varphi=\varphi(x,t,u)$
\cite{ib-2011}. Thus, by Theorem \ref{th-4}, any perturbed equation
associated with Eq.(\ref{short})
\begin{equation}\label{shortper}
\no u_{xt}=u+\frac{1}{6}(u^3)_{xx}+\epsilon f(x,t,u,u_x,u_t,\dots),
\end{equation}
with arbitrary function $f(x,t,u,u_x,u_t,\dots)$, is not
approximately nonlinearly self-adjoint in the sense of Definition
\ref{def-1}.

In what follows, based on Theorem \ref{con-law}, we give a specific
approximate conservation law formula by means of the known
conservation law of unperturbed system (\ref{perturb}), which
simplifies the computations of first-order approximate conservation
laws for perturbed system (\ref{perturbnew}).

\begin{theorem}\label{app-law}
Let $X = X_0 +\epsilon X_1$ be a first-order approximate symmetry
generator of perturbed system (\ref{perturbnew}), where $X_0
=\xi_0^i\partial_{x^i} + \eta_0^{\sigma}\partial_{u^{\sigma}},X_1 =
\xi_1^i\partial_{x^i} + \eta_1^{\sigma}\partial_{u^{\sigma}}$ and
the substitution for approximate nonlinear self-adjointness is
$w=w_0+\epsilon w_1=\psi(x,u)+\epsilon \phi(x,u)$. If
$C=(C^1,\dots,C^n)$ is a conserved vector of unperturbed system
(\ref{perturb}) obtained by formula (\ref{formula}) where the term
$\xi^i\mathcal {K}$ is not omitted, then a first-order approximate
conservation law $D_i(T^i)=O(\epsilon^2)$ of perturbed system
(\ref{perturbnew}) is given by the formula
%$T=(T^1,\dots,T^n)$
\begin{eqnarray}\label{approximate-law}
&&\no\hspace{-0.6cm}T^i=C^i_{|_{v=w}}+\epsilon\Big[\xi_0^i\mathcal
{L}_1+W_0^{\sigma}\frac{\delta \mathcal {L}_1}{\delta
u_i^{\sigma}}+D_j(W_0^{\sigma})\frac{\delta \mathcal {L}_1}{\delta
u_{ij}^{\sigma}}+ D_jD_k(W_0^{\sigma})\frac{\delta \mathcal
{L}_1}{\delta u_{ijk}^{\sigma}}+
\dots\\
&&\hspace{2.1cm}+ \xi_1^i\mathcal
{L}_2+W_1^{\sigma}\frac{\delta\mathcal {L}_2}{\delta
u_i^{\sigma}}+D_j(W_1^{\sigma})\frac{\delta \mathcal {L}_2}{\delta
u_{ij}^{\sigma}}+ D_jD_k(W_1^{\sigma})\frac{\delta \mathcal
{L}_2}{\delta u_{ijk}^{\sigma}}+\dots\Big]_{|_{w_0=\psi}},
\end{eqnarray}
where $W_0^{\sigma}=\eta_0^{\sigma}-\xi_0^ju^{\sigma}_{j}$,
$W_1^{\sigma}=\eta_1^{\sigma}-\xi_1^ju^{\sigma}_{j}$ and $\mathcal
{L}_1=w_0^{\beta}E_{\beta}^1,\mathcal {L}_2=w_0^\beta E_{\beta}^0$.
The Euler-Lagrange operators with respect to derivatives of
$u^\sigma$ are obtained from (\ref{var-op}) by replacing $u^\sigma$
by the corresponding derivatives, e.g.
 \begin{eqnarray}
\no\frac{\delta}{\delta u_i^{\sigma}}=\frac{\partial}{\partial
u_i^{\sigma}}+\sum_{s=1}^{\infty}(-1)^sD_{i_1}\dots
D_{i_s}\frac{\partial}{\partial u^{\sigma}_{ii_1\dots i_s}}.
\end{eqnarray}
\end{theorem}

\emph{Proof.} Following the symbols in Theorem \ref{app-law}, that
means
\begin{eqnarray}
&&\no \xi^i=\xi_0^i+\epsilon\xi_1^i,~~W^{\sigma}=W_0^{\sigma}+\epsilon W_1^{\sigma}, \\
&&\no \mathcal {K}=(w_0^{\beta}+\epsilon w_1^{\beta})(
E_{\beta}^0+\epsilon
E_{\beta}^1)=w_0^{\beta}E_{\beta}^0+\epsilon(w_0^{\beta}E_{\beta}^1+w_1^{\beta}E_{\beta}^0)+\epsilon^2w_1^{\beta}E_{\beta}^1,
\end{eqnarray}
then inserting them into formula (\ref{formula}), we have
\begin{eqnarray}
&&\no T^i=(\xi_0^i+\epsilon\xi_1^i)[w_0^{\beta}E_{\beta}^0
+\epsilon(w_0^{\beta}E_{\beta}^1+w_1^{\beta}E_{\beta}^0)]+(W_0^{\sigma}+\epsilon
W_1^{\sigma})\frac{\delta [w_0^{\beta}E_{\beta}^0
+\epsilon(w_0^{\beta}E_{\beta}^1+w_1^{\beta}E_{\beta}^0)]}{\delta
u_i^{\sigma}}\\\no
&&\hspace{1.2cm} +D_j(W_0^{\sigma}+\epsilon W_1^{\sigma})
\frac{\delta [w_0^{\beta}E_{\beta}^0
+\epsilon(w_0^{\beta}E_{\beta}^1+w_1^{\beta}E_{\beta}^0)]}{\delta
u_{ij}^{\sigma}}\\\no
&&\hspace{1.2cm}+D_jD_k(W_0^{\sigma}+\epsilon W_1^{\sigma})
\frac{\delta [w_0^{\beta}E_{\beta}^0
+\epsilon(w_0^{\beta}E_{\beta}^1+w_1^{\beta}E_{\beta}^0)]}{\delta
u_{ijk}^{\sigma}
}+\dots\\\no&&\hspace{0.5cm}=\xi_0^i(w_0^{\beta}+\epsilon
w_1^{\beta})E_{\beta}^0+W_0^{\sigma}\frac{\delta
[(w_0^{\beta}+\epsilon w_1^{\beta})E_{\beta}^0]}{\delta
u_i^{\sigma}}+D_j(W_0^{\sigma}) \frac{\delta [(w_0^{\beta}+\epsilon
w_1^{\beta})E_{\beta}^0]}{\delta u_{ij}^{\sigma}}+\dots\\\no
&&\hspace{1.2cm}+\epsilon\Big[\xi_0^iw_0^{\beta}E_{\beta}^1+W_0^{\sigma}\frac{\delta
(w_0^{\beta}E_{\beta}^1 )}{\delta
u_i^{\sigma}}+D_j(W_0^{\sigma})\frac{\delta (w_0^{\beta}E_{\beta}^1
)}{\delta u_{ij}^{\sigma}}+\dots
\\\no
&&\hspace{1.2cm}+\xi_1^iw_0^{\beta}E_{\beta}^0+W_1^{\sigma}\frac{\delta
(w_0^{\beta}E_{\beta}^0 )}{\delta
u_i^{\sigma}}+D_j(W_0^{\sigma})\frac{\delta (w_0^{\beta}E_{\beta}^0
)}{\delta u_{ij}^{\sigma}}+\dots\Big]\\\no
&&\hspace{0.5cm}=C^i_{|_{v=w}}+\epsilon\Big[\xi_0^i\mathcal
{L}_1+W_0^{\sigma}\frac{\delta \mathcal {L}_1}{\delta
u_i^{\sigma}}+D_j(W_0^{\sigma})\frac{\delta \mathcal {L}_1}{\delta
u_{ij}^{\sigma}}+ D_jD_k(W_0^{\sigma})\frac{\delta \mathcal
{L}_1}{\delta u_{ijk}^{\sigma}}+
\dots\\
&&\no\hspace{1.2cm}+ \xi_1^i\mathcal
{L}_2+W_1^{\sigma}\frac{\delta\mathcal {L}_2}{\delta
u_i^{\sigma}}+D_j(W_1^{\sigma})\frac{\delta \mathcal {L}_2}{\delta
u_{ij}^{\sigma}}+ D_jD_k(W_1^{\sigma})\frac{\delta \mathcal
{L}_2}{\delta u_{ijk}^{\sigma}}+\dots\Big]_{|_{w_0=\psi}},
\end{eqnarray}
where $\mathcal {L}_1=w_0^{\beta}E_{\beta}^1, \mathcal
{L}_2=w_0^{\beta}E_{\beta}^0$ and the second-order terms of
$\epsilon$ are omitted, thus completes the proof. $\hfill{}
\blacksquare$

Theorem \ref{app-law} proposes a direct way to construct approximate
conservation laws based on the known conservation laws of the
corresponding unperturbed PDEs. %, especially effective for the perturbed system
%consisting of two or more perturbed pdes and sharing the same
%unperturbed system.
It means that if unperturbed system (\ref{perturb}) has a
conservation law written as (\ref{formula}), then formula
(\ref{approximate-law}) in Theorem \ref{app-law} gives a direct
computable formula of first-order approximate conservation laws for
perturbed system (\ref{perturbnew}). Furthermore, the approach of
approximate symmetry definitely extends symmetry scope of perturbed
PDEs \cite{ba-1991,fs-1989}, thus approximate nonlinear
self-adjointness can generate new approximate conservation laws
which cannot be obtain by nonlinear self-adjointness of PDEs.
\begin{remark}
In Theorem \ref{app-law}, the conservation law $D_i(C^i)=0$ is
calculated via the substitution $v=\varphi(x,u)$ on the solution
manifold of system (\ref{perturb}). The approximate conservation law
$D_i(T^i)=O(\epsilon^2)$ is obtained by formula
(\ref{approximate-law}) evaluated on the solution manifold of
perturbed system (\ref{perturbnew}).

Generally speaking, the term $\epsilon\,\xi_1^i\mathcal {L}_2$ can
be omitted because it becomes second-order of $\epsilon$ on the
solution manifold of system (\ref{perturbnew}).
%
%In , the conserved vector $C=(C^1,\dots,C^n)$ contains the
%introduced variable $v$ while approximate conserved vector
%$T=(T^1,\dots,T^n)$ involves $v$ and $w$, thus by Corollary
%\ref{cor-1}, the substitutions , which will be used to construct
%local first-order approximate conservation laws.
\end{remark}

\section{Applications}
We illustrate our results by considering a class of perturbed wave
equations
\begin{eqnarray}\label{wave-ger}
u_{tt}-[F(u)u_x]_x+\epsilon G(u)u_t=0,
\end{eqnarray}
where $F(u),G(u)$ are two arbitrary smooth functions.
Eq.(\ref{wave-ger}) arises in the applied problems such as wave
phenomena in shallow water, long radio engineering lines and
isentropic motion of a fluid in a pipe, etc. \cite{ba-1991,char}. In
\cite{zhang1,ba-1991} and reference therein, approximate symmetry
classification and reductions were performed by two approximate
symmetry methods. Approximate conditional symmetries and approximate
conditional invariant solutions for $F(u)=u$ was considered in
\cite{qu1}. We discussed approximate conservation laws by means of
approximate nonlinear self-adjointness for $G(u)=1$ in
\cite{zhang-2013} .

By Theorem \ref{th-3}, we directly have the following proposition.
\begin{prop}\label{pro-1}
Eq.(\ref{wave-ger}) is approximately  nonlinearly self-adjoint for
any functions $F(u)$ and $G(u)$.
\end{prop}

\emph{Proof:} In \cite{ib-2011}, the authors showed that the
unperturbed equation of Eq.(\ref{wave-ger}) with arbitrary function
$F(u)$
\begin{eqnarray}\label{unperturbed}
&&u_{tt}-(F(u)u_x)_x=0,
\end{eqnarray}
is nonlinearly self-adjoint upon the substitution
$v=c_1tx+c_2x+c_3t+c_4$ with arbitrary parameter
$c_i\,(i=1,\dots,4)$ such that $v\neq0$, thus by Theorem \ref{th-5},
Eq.(\ref{wave-ger}) is approximately nonlinearly self-adjoint. This
completes the proof. $\hfill{}\blacksquare$

%By proposition \ref{pro-1}, we can use identity (\ref{forper}) to
%find special expression of substitution for special expression of
%$F(u)$ and $G(u)$.
In what follows, we first classify all possible cases of
Eq.(\ref{wave-ger}) to be approximate nonlinear self-adjointness,
and then use formula (\ref{approximate-law}) to construct
approximate conservation laws for some special cases. Note that in
this section $c_i\,(i=1,\dots,8)$ are arbitrary constants such that
the substitution $w\neq0$.

\subsection{Classification for approximate nonlinear self-adjointness}
Let the formal Lagrangian of Eq.(\ref{wave-ger})
\begin{eqnarray}
&&\no\mathcal{K}=w\left[u_{tt}-F'(u)u_x^2-F(u)u_{xx}+\epsilon G(u)
u_t\right],
\end{eqnarray}
then the adjoint equation is $\delta \mathcal {K}/\delta
u=w_{tt}-F(u)w_{xx}-\epsilon\, G(u)w_t=0$. Assume the substitution
of approximate nonlinear self-adjointness is $w=\psi(x,t,u)+\epsilon
\phi(x,t,u)$, then by equivalent equality (\ref{forper}), we have
\begin{eqnarray}
\no\left[w_{tt}-F(u)w_{xx}-\epsilon
G(u)w_t\right]_{|\{w=\psi+\epsilon
\phi\}}=(\lambda+\epsilon\mu)[u_{tt}-(F(u)u_x)_x]+\epsilon\lambda
G(u) u_t,
\end{eqnarray}
with undetermined functions $\lambda(x,t,u)$ and $\mu(x,t,u)$, thus
after proper arrangements, $\psi$ and $\phi$ satisfy
\begin{eqnarray}\label{deter}
&&\no \lambda F'(u)=\mu F'(u)=0,\\
&&\no \psi_{uu}=\psi_{tu}=\psi_{xu}=0,\\\no
&&\phi_u-\mu=0,\phi_{uu}=\phi_{xu}=0,\\\no
&&\psi_u-\lambda=0,\psi_{tt}-F(u)\psi_{xx}=0,\\
&&\phi_{tu}-\lambda G(u)=0,\phi_{tt}-F(u)\phi_{xx}-G(u)\psi_t=0.
\end{eqnarray}

The required substitution for approximate nonlinear self-adjointness
is to solve system (\ref{deter}), thus we consider all possible
cases about $F(u)$ and $G(u)$ which make system (\ref{deter})
possess nontrivial solutions.

For any arbitrary functions $F(u)$ and $G(u)$, the substitution for
approximate nonlinear self-adjointness is given in the following
proposition.
\begin{prop}
For any arbitrary functions $F(u)$ and $G(u)$, Eq.(\ref{wave-ger})
is approximately nonlinearly self-adjoint via a substitution
$w=c_1x+c_2+\epsilon (c_3tx+c_4x+c_5t+c_6)$.
\end{prop}

For some special functions $F(u)$ and $G(u)$, more affluent
substitutions are found. Thus from the first two equations in system
(\ref{deter}), two different cases about $F(u)$ are considered.
\subsubsection{$F'(u)=0$}

Condition $F'(u)=0$ means that $F(u)$ is a constant, without loss of
generality, set $F(u)=1$, thus Eq.(\ref{wave-ger}) becomes
\begin{equation}\label{wave-per-1}
u_{tt}-u_{xx}+\epsilon G(u)u_t=0.
\end{equation}

In this case, solving system (\ref{deter}) gives
$\psi=c_1u+f(x,t),\phi=\mu(t)u+g(x,t)$ and
\begin{eqnarray}\label{equ-1}
&&\no  \mu_t-c_1 G(u)=0,\\
&& \mu_{tt}u+g_{tt}-G(u) f_{tt}-g_{xx}=0,
\end{eqnarray}
%$ \varphi=-c_1u+h(x,t)$, where $f$ satisfies $f_{tt}-f_{xx}=0$,
%$g(x,t)$ and $h(x,t)$ solve $g_{tt}-g_{xx}-G(u)h_t=0$,
thus two different cases arise.

\textbf{Case 1.1} $G(u)$ is a constant.

Let $G(u)=1$, Eq.(\ref{wave-per-1}) becomes a linear perturbed
equation in the form
\begin{equation}\label{wave-per-11}
u_{tt}-u_{xx}+\epsilon u_t=0,
\end{equation}
whose approximate conservation laws are studied in
\cite{ah,ah-1999}.

It is easy to obtain the following results for linear perturbed
equation (\ref{wave-per-11}) based on the above computations.
\begin{prop}\label{prop-2}
Eq.(\ref{wave-per-11}) is approximately nonlinearly self-adjoint via
a substitution $w=c_1u+f(x,t)+\epsilon[c_1tu+c_2u+g(x,t)]$, where
$f(x,t)$ and $g(x,t)$ solve $u_{tt}-u_{xx}=0$.
\end{prop}

\textbf{Case 1.2} $G(u)$ is not a constant.

In this case, since $G(u)$ is a nonconstant function, thus
$\mu_t=c_1=0$, substituting it into the second equation in
(\ref{equ-1}), we have the following results.
\begin{prop}\label{prop-3}
Eq.(\ref{wave-per-1}) is approximately nonlinearly self-adjoint via
a substitution
\begin{equation}
w=f(x)t+g(x)+\epsilon[c_2u+h(x,t)],
 \end{equation}
where $h(x,t)$ satisfy $h_{tt}=h_{xx}$ and $f(x),g(x)$  are
arbitrary functions.
\end{prop}

\subsubsection{$F'(u)\neq0$}
Since $F(u)$ is a nonconstant function, thus $\lambda=\mu=0$ and
system (\ref{deter}) becomes
\begin{eqnarray}\label{deter-1}
&&\no
\psi_u=\phi_u=0,\\&&\no\psi_{tt}=\psi_{xx}=0,\\
&&\phi_{tt}-F(u)\phi_{xx}-G(u)\psi_t=0.
\end{eqnarray}
Further discussion of system (\ref{deter-1}) needs two different
cases to be considered.

\textbf{Case 2.1} $G(u)$ is a constant.

Set $G(u)=1$, then Eq.(\ref{wave-ger}) becomes
\begin{eqnarray}\label{wave}
u_{tt}+\epsilon u_t=[F(u)u_x]_x.
\end{eqnarray}

In \cite{zhang-2013}, we have showed that Eq.(\ref{wave}) is
approximately nonlinearly self-adjoint, thus we recall the results
in order to make the classification complete.
\begin{prop}\label{prop-1}
Eq.(\ref{wave}) is approximately nonlinearly self-adjoint under the
substitution
\begin{eqnarray}\label{25}
w=c_1tx+c_2x+c_3t+c_4+\epsilon(\frac{1}{2}c_1t^2x+c_5tx+c_6x+\frac{1}{2}c_3t^2+c_7t+c_8).
\end{eqnarray}
\end{prop}

\textbf{Case 2.2} $G(u)$ is not a constant.

Nonconstant function $G(u)$ implies that the last equation in system
(\ref{deter-1}) is divided into two equations
$\phi_{tt}=0,F(u)\phi_{xx}+G(u)\psi_t=0$, thus two different cases
come into being. If $F(u)\neq G(u)$, no additional substitution are
found, but for $F(u)=G(u)$, new substitution is listed as follows.

\begin{prop}\label{prop-4}
 If $F(u)=G(u)$, then Eq.(\ref{wave-ger}) is approximately
nonlinearly self-adjoint under the substitution
\begin{eqnarray}\label{wave-fg}
w=c_1tx+c_2x+c_3t+c_4+\epsilon\big(c_5xt+c_6t-\frac{1}{6}c_1x^3-\frac{1}{2}c_3x^2+c_7x+c_8\big).
\end{eqnarray}
%
%
%2. If $F(u)\neq G(u)$, then Eq.(\ref{wave-ger}) is approximately
%nonlinearly self-adjoint under the substitution
%\begin{eqnarray}
%w=c_1x+c_2+\epsilon(c_3tx+c_4x+c_5t+c_6).
%\end{eqnarray}
\end{prop}

\subsection{Approximate conservation law}
In this subsection, we will use the above classification results for
approximate nonlinear self-adjointness and formula
(\ref{approximate-law}) to construct approximate conservation law
formula of Eq.(\ref{wave-ger}) which is given as follows.
%
%With the help of formula (\ref{approximate-law}), the general
%approximate conservation law formula for Eq.(\ref{wave-ger}) is

\begin{prop}
The general approximate conservation law formula for
Eq.(\ref{wave-ger}) is expressed by
\begin{eqnarray}\label{formula-wave}
&&
T_{(\ref{per-1})}^t=(C_{(\ref{unperturbed})}^t)_{|_{v=w}}+\epsilon\left[\xi_0^twG(u)u_t+wG(u)
W_0-w_t W_1+w D_t(W_1)\right],
\\\no &&T_{(\ref{per-1})}^x=(C_{(\ref{unperturbed})}^x)_{|_{v=w}}+\epsilon\left[\xi_0^x w G(u)u_t+W_1(F(u)w_x-F'(u)u_xw)-w
F(u)D_x(W_1)\right],
\end{eqnarray}
where the conserved densities $C_{(\ref{unperturbed})}^t$ and
$C_{(\ref{unperturbed})}^x$ for equation (\ref{unperturbed}) is
given by
\begin{eqnarray}\label{formula-wave-1}
&&\no C_{(\ref{unperturbed})}^t=\xi_0^tv[u_{tt}-(F(u)u_x)_x]-v_t
W_0+v D_t(W_0),
\\&& C_{(\ref{unperturbed})}^x=\xi_0^x v[u_{tt}-(F(u)u_x)_x]+W_0(F(u)v_x-F'(u)u_xv)-v F(u)
D_x(W_0),
\end{eqnarray}
where $v$ is the substitution for nonlinear self-adjointness of
Eq.(\ref{unperturbed}). The first-order approximate infinitesimal
operator $X=X_0+\epsilon
X_1=X_0+\epsilon(\xi_1^t\partial_t+\xi_1^x\partial_x+\eta_1\partial_u)$
and the infinitesimal operator
$X_0=\xi_0^t\partial_t+\xi_0^x\partial_x+\eta_0\partial_u$ leaves
Eq.(\ref{unperturbed}) invariant and
$W_0=\eta_0-\xi_0^xu_x-\xi_0^tu_t$,
$W_1=\eta_1-\xi_1^xu_x-\xi_1^tu_t$.
\end{prop}

Given an approximate symmetry of Eq.(\ref{wave-ger}), then one can
use the general formula (\ref{formula-wave}) to construct
approximate conservation laws based on the conservation laws
(\ref{formula-wave-1}) of unperturbed equation (\ref{unperturbed}).
Note that the approximate symmetry classification of
Eq.(\ref{wave-ger}) is performed in \cite{ba-1991}. Thus in what
follows, we consider several special examples.
\subsubsection{Two Single perturbed PDEs}
We first consider two single special perturbed PDEs of equation
(\ref{wave-ger}).

\textbf{Example 1.} The first consideration is to search for
approximate conservation laws of Eq.(\ref{wave}). The first step is
to construct conservation laws of unperturbed Eq.(\ref{unperturbed})
which are expressed by formula (\ref{formula-wave-1}).  Then by
formula (\ref{formula-wave}), we obtain first-order approximate
conservation law of Eq.(\ref{wave}) in the form
\begin{eqnarray}\label{con-for-wave}
&&\no
T_{(\ref{wave})}^t=(C_{(\ref{unperturbed})}^t)_{|_{v=w}}+\epsilon\left[\xi_0^twu_t+w
W_0-w_t W_1+w D_t(W_1)\right],
\\ &&T_{(\ref{wave})}^x=(C_{(\ref{unperturbed})}^x)_{|_{v=w}}+\epsilon\big[\xi_0^x wu_t+W_1(F(u)w_x-F'(u)u_xw)-w
F(u)D_x(W_1)\big],
\end{eqnarray}
where $w$ is given by (\ref{25}).

Consider a special operator $X=x\partial_x+t\partial_t+\epsilon
(\frac{1}{2}t^2\partial_t-2t\partial_u)$ when $F(u)=e^u$, then
$\xi_0^t=t,\xi_0^x=x,W_0=-xu_x-tu_t$ and
$\xi_1^t=\frac{1}{2}t^2,\xi_1^x=0,W_1=-2t-\frac{1}{2}t^2u_t$, thus
by formula (\ref{con-for-wave}), we find
\begin{eqnarray}
&&\no
T_{(\ref{wave})}^t=te^uw_xu_x+xu_xw_t+tu_tw_t+xw_xu_t\\\no&&\hspace{1.5cm}+\epsilon\big(2tw_t+\frac{1}{2}t^2u_tw_t-w
xu_x+\frac{1}{2}t^2e^uw_xu_x-twu_t-2w\big),
\\\no &&T_{(\ref{wave})}^x=-xw_tu_{t}-xe^uu_xw_x-te^uu_tw_x-te^uw_tu_x\\\no
&&\hspace{1.5cm}+\epsilon\big(x
wu_t-\frac{1}{2}t^2e^uw_tu_x-2te^uw_x-\frac{1}{2}t^2e^uu_tw_x+te^uu_xw\big),
\end{eqnarray}
which makes
$D_t(T_{(\ref{wave})}^t)+D_x(T_{(\ref{wave})}^x)=\frac{1}{4}\epsilon^2
\Omega $ on the  solution manifold of Eq.(\ref{wave}), where
$\Omega$ is a large expression and thus is omitted. In particular,
let $w=x$, then
\begin{eqnarray}
&&\no
(T_{(\ref{wave})}^t)_{|w=x}=te^uu_x+xu_t+\epsilon\big(\frac{1}{2}t^2e^uu_x-
x^2u_x-xtu_t-2x\big),
\\\no
&&(T_{(\ref{wave})}^x)_{|w=x}=-xe^uu_x-te^uu_t+\epsilon\big(x^2
u_t-2te^u-\frac{1}{2}t^2e^uu_t+txe^uu_x\big),
\end{eqnarray}
which is identical to the results in \cite{zhang-2013}.

\textbf{Example 2.}  The second example is to consider a perturbed
nonlinear wave equation
\begin{equation}\label{per-example2}
u_{tt}-(uu_{x})_x+\epsilon uu_t=0.
\end{equation}

Similarly, using formula (\ref{formula-wave}), we obtain
\begin{eqnarray}\label{formula-wave-3}
&&\no
T_{(\ref{per-example2})}^t=(C_{(\ref{unperturbed})}^t)_{|_{\{v=w,F(u)=u\}}}+\epsilon\left[\xi_0^twuu_t+wu
W_0-w_t W_1+w D_t(W_1)\right],
\\ &&T_{(\ref{per-example2})}^x=(C_{(\ref{unperturbed})}^x)_{|_{\{v=w,F(u)=u\}}}+\epsilon\left[\xi_0^x wuu_t+W_1(uw_x-u_xw)-w
uD_x(W_1)\right],
\end{eqnarray}
where the substitution $w$ is given by (\ref{wave-fg}) in
Proposition \ref{prop-4}.

In particular, consider an approximate symmetry
$X=t\partial_t+\frac{1}{2}x\partial_x-u\partial_u$, then
$\xi_0^t=t$,
$\xi_0^x=\frac{1}{2}x,W_0=-u-\frac{1}{2}xu_x-tu_t,\xi_1^t=\xi_1^x=0,W_1=0$,
thus an approximate conservation law with the conserved density is
given by
\begin{eqnarray}\label{formula-wave-4}
&&\no T_{(\ref{per-example2})}^t=w_t
(u+\frac{1}{2}xu_x+tu_t)+\frac{1}{2}xw_xu_t+tw_xuu_{x}-\frac{3}{2}wu_t-\epsilon
w u(u+\frac{1}{2}xu_x),\\
\no&&
T_{(\ref{per-example2})}^x=-uw_x(u+\frac{1}{2}xu_x+tu_t)-\frac{1}{2}xw_tu_{t}-tw_tuu_x+\frac{3}{2}wuu_x+\frac{1}{2}\epsilon
x w uu_t,
\end{eqnarray}
which makes
\begin{equation}
\no
D_t(T_{(\ref{per-example2})}^t)+D_x(T_{(\ref{per-example2})}^x)=-\frac{1}{2}\epsilon^2(c_5x+c_6)(2u+2tu_t+xu_x)
\end{equation}
on the  solution manifold of Eq.(\ref{per-example2}). Especially,
set $w=t-\frac{1}{2}\epsilon x^2$, one obtains
\begin{eqnarray}\label{formula-wave-5}
&&\no
T_{(\ref{per-example2})}^t=u+\frac{1}{2}xu_x-\frac{1}{2}tu_t-\frac{1}{2}x^2u_t-xtuu_{x}-\epsilon
t u(u+\frac{1}{2}xu_x-\frac{3}{4}
x^2u_t),\\
\no&&
T_{(\ref{per-example2})}^x=xu^2+\frac{1}{2}x^2uu_x+xtuu_t-\frac{1}{2}xu_{t}+\frac{1}{2}tuu_x-\epsilon
xu (\frac{3}{4}xu_x-\frac{1}{2}tu_t),
\end{eqnarray}
which makes $D_t(T^t)+D_x(T^x)=0$.
\subsubsection{Two perturbed wave equations having the same
unperturbed equation}

Consider the two perturbed wave equations with the form
\begin{eqnarray}\label{per-1}
u_{tt}-u_{xx}+\epsilon u_t=0,
\end{eqnarray}
and
\begin{eqnarray}\label{per-2}
u_{tt}-u_{xx}+\epsilon(uu_t+\frac{1}{2}tu_t^2-\frac{1}{2}tu_x^2)=0,
\end{eqnarray}
whose approximate conservation laws are studied in \cite{ah-1999,ah}
by partial Lagrangian method. The above two perturbed equations
share the same unperturbed equation
\begin{eqnarray}\label{nonper}
u_{tt}-u_{xx}=0,
\end{eqnarray}
which is the classical wave equation and describes undamped linear
waves in an isotropic medium. It is easy to show that
Eq.(\ref{nonper}) is nonlinearly self-adjoint via a substitution
$v=c_1u+f(x,t)$, where $f$ satisfies $f_{tt}-f_{xx}=0$. Since
Eq.(\ref{nonper}) is nonlinearly self-adjoint, thus by Theorem
\ref{th-3}, both Eq.(\ref{per-1}) and Eq.(\ref{per-2}) are
approximately nonlinearly self-adjoint. Specifically, we have the
following results.
\begin{prop}
 Eq.(\ref{per-2}) is approximately nonlinearly self-adjoint upon a
substitution $w=c_1u+f+\epsilon(\frac{3}{4}c_1tu^2+\frac{1}{2}t f
u+g)$ with functions $f(x,t)$ and $g(x,t)$ satisfying
Eq.(\ref{nonper}).
\end{prop}

\emph{Proof:} Let the formal Lagrangian $\mathcal
{L}=w\big(u_{tt}-u_{xx}+\epsilon(uu_t+\frac{1}{2}tu_t^2-\frac{1}{2}tu_x^2)\big)$,
then the adjoint equation is $\delta\mathcal {L}/\delta
u=w_{tt}-w_{xx}-\epsilon[(uw)_t+t(u_tw)_t-t(u_xw)_x]$. Assume that
the substitution is in the form $w=\psi(x,t,u)+\epsilon\phi(x,t,u)$
and should satisfy equality (\ref{forper}), in this case, it becomes
\begin{eqnarray}
&&\no \big(w_{tt}-w_{xx}-\epsilon (u w)_t-\epsilon
t(u_tw)_t+\epsilon
t(u_xw)_x\big)_{|_{w=\psi(x,t,u)+\epsilon\phi(x,t,u)}}
\\&&\hspace{4cm}=(\lambda_0+\epsilon
\mu_0)(u_{tt}-u_{xx})+\epsilon\lambda_0\big(uu_t+\frac{1}{2}tu_t^2-\frac{1}{2}tu_x^2\big),
\end{eqnarray}
where we omit the terms containing $\epsilon^2$.

The reckoning shows that $ \psi(x,t,u),\phi(x,t,u)$ and the
undetermined functions $\lambda_0,\mu_0$ satisfy
\begin{eqnarray}\label{equ}
&&\no \psi_{uu}=\psi_{xu}=\psi_{tu}=0,\\&&\no \psi_u=\lambda_0,
\psi_{tt}-\psi_{xx}=0,\\&&\no \phi_u-\mu_0-t\psi=0,
~2\phi_{uu}-2t\psi_u-\lambda_0 t=0,\\&&
2\phi_{tu}-\psi_uu+\psi+t\psi_t-\lambda_0u=0,~
2\phi_{xu}-t\psi_x=0,~\phi_{tt}-\phi_{xx}-u\psi_t=0.
\end{eqnarray}
Solving system (\ref{equ}) gives
$\psi=c_1u+f,\phi=\frac{3}{4}c_1tu^2+\frac{1}{2}t f u+g$ and
$\lambda_0=c_1,\mu_0=\frac{1}{2}c_1tu-\frac{1}{2}tf$. It proves the
result. $\hfill{} \blacksquare$

Next, we construct approximate conservation laws by means of Theorem
\ref{app-law}. Firstly, with the help of formula
(\ref{formula-wave-1}), we obtain a conservation law formula of
Eq.(\ref{nonper})
\begin{eqnarray}\label{un-con}
&&\no C^t_{(\ref{nonper})}=\xi_0^tv(u_{tt}-u_{xx})-v_t W_0+v
D_t(W_0),
\\&& C^x_{(\ref{nonper})}=\xi_0^x v(u_{tt}-u_{xx})+v_x W_0-v
D_x(W_0).
\end{eqnarray}

The approximate conservation laws of equations (\ref{per-1}) and
(\ref{per-2}) can be obtained by means of formula
(\ref{formula-wave}) via expression (\ref{un-con}).

\textbf{Approximate conservation law of Eq.(\ref{per-1}):} By means
of the substitution $w=c_1u+h+\epsilon(c_1tu+c_2u+k)$ and formula
(\ref{approximate-law}), we obtain first-order approximate
conservation law formula of Eq.(\ref{per-1})
\begin{eqnarray}\label{con-for-1}
&&\no
T_{(\ref{per-1})}^t=(C^t_{(\ref{nonper})})_{|_{v=w}}+\epsilon\left(\xi_0^twu_t+w
W_0-w_t W_1+w D_t(W_1)\right),
\\&&T_{(\ref{per-1})}^x=(C^x_{(\ref{nonper})})_{|_{v=w}}+\epsilon\big(\xi_0^x wu_t+w_x
W_1-w D_x(W_1)\big),
\end{eqnarray}
with the first-order approximate infinitesimal operator
$X=X_0+\epsilon
X_1=X_0+\epsilon(\xi_1^t\partial_t+\xi_1^x\partial_x+\eta_1\partial_u)$
admitted by Eq.(\ref{per-1}) and $W_1=\eta_1-\xi_1^xu_x-\xi_1^tu_t$.

In particular, consider an approximate symmetry
$X=\partial_t+\epsilon(-\frac{1}{2}u\partial_u)$, then
$\xi_0^t=1,\xi_0^x=0,W_0=-u_t,W_1=-\frac{1}{2}u$, thus an
approximate conservation law  is
\begin{eqnarray}\label{con-for-11}
&&\no T_{(\ref{per-1})}^t=u_tw_t+w_xu_x+\epsilon\big(\frac{1}{2}uw_t
-\frac{1}{2}u_tw \big),
\\&&T_{(\ref{per-1})}^x=-u_tw_x-w_tu_x+\epsilon\big(\frac{1}{2}u_xw-\frac{1}{2}uw_x
\big).
\end{eqnarray}

Induced by the approximate operator
$X=\partial_t+\epsilon(-\frac{1}{2}u\partial_u)$, formula
(\ref{con-for-11}) provides more approximate conservation laws via
substitution $w=c_1u+f+\epsilon(c_1tu+c_2u+g)$ which includes the
obtained approximate conservation laws in \cite{ah}. In particular,
let $w=\frac{1}{2}u+\epsilon\frac{1}{2}tu$, then after proper
arrangements, conservation law (\ref{con-for-11}) becomes
\begin{eqnarray}
&&\no
T_{(\ref{per-1})}^t=\frac{1}{2}u_t^2+\frac{1}{2}u_x^2+\frac{1}{2}\epsilon\big(
tu^2_t+uu_t+tu_x^2\big),\\\no
&&T_{(\ref{per-1})}^x=-u_xu_t-\epsilon\big(\frac{1}{2}uu_x+tu_xu_t\big),
\end{eqnarray}
which coincides with the results in \cite{ah} up to some so-called
gauge terms stated there.

\textbf{Approximate conservation law of Eq.(\ref{per-2}): } For
Eq.(\ref{per-2}), using the substitution
$w=c_1u+f+\epsilon(\frac{3}{4}c_1tu^2+\frac{1}{2}t f u+g)$ with
$f,g$ satisfying Eq.(\ref{nonper}),  one has
\begin{eqnarray}\label{con-for-2}
&&\no\hspace{-0.2cm}
T_{(\ref{per-2})}^t=(C^t_{(\ref{nonper})})_{|_{v=w}}+\epsilon\Big(\xi_0^t(uu_t+\frac{1}{2}tu_t^2-\frac{1}{2}tu_x^2)w+
(u+tu_t)wW_0-w_t \widetilde{W}_1+w D_t(\widetilde{W}_1)\Big),
\\&&\hspace{-0.2cm}T_{(\ref{per-2})}^x=(C^x_{(\ref{nonper})})_{|_{v=w}}
+\epsilon\Big(\xi_0^x
(uu_t+\frac{1}{2}tu_t^2-\frac{1}{2}tu_x^2)w-tu_xwW_0+w_x
\widetilde{W}_1-w D_x(\widetilde{W}_1)\Big),
\end{eqnarray}
with the first-order approximate infinitesimal operator
$\widetilde{X}=X_0+\epsilon
\widetilde{X}_1=X_0+\epsilon(\widetilde{\xi}_1^t\partial_t+\widetilde{\xi}_1^x\partial_x+\widetilde{\eta}_1\partial_u)$
admitted by Eq.(\ref{per-2}) and
$\widetilde{W}_1=\widetilde{\eta}_1-\widetilde{\xi}_1^xu_x-\widetilde{\xi}_1^tu_t$.

For example, consider an approximate Lie point symmetry
$\widetilde{X}=\partial_u+\epsilon(-\frac{1}{2}tu\partial_u)$, it
means $\xi_0^t=\xi_0^x=0,W_0=1,\widetilde{W}_1=-\frac{1}{2}tu$, then
by (\ref{un-con}) and (\ref{con-for-2}), we obtain
\begin{eqnarray}\label{con-for-21}
&& \no T_{(\ref{per-2})}^t=-w_t+\epsilon\big(
\frac{1}{2}tw_tu+\frac{1}{2}w u+\frac{1}{2}t w u_t\big),\\
&&
T_{(\ref{per-2})}^x=w_x-\epsilon\big(\frac{1}{2}twu_x+\frac{1}{2}tuw_x
\big).
\end{eqnarray}

Similarly,  with substitution
$w=c_1u+f+\epsilon(\frac{3}{4}c_1tu^2+\frac{1}{2}t f u+g)$, formula
(\ref{con-for-21}) also gives more approximate conservation laws
generated by the approximate operator
$\widetilde{X}=\partial_u+\epsilon(-\frac{1}{2}tu\partial_u)$. Let
$w=u+\frac{3}{4}\epsilon tu^2$, then (\ref{con-for-21}) becomes
\begin{eqnarray}
&&\no T_{(\ref{per-2})}^t=-u_t-\epsilon\big(
\frac{1}{2}tuu_t+\frac{1}{4}u^2\big),~~T_{(\ref{per-2})}^x=u_x+\frac{1}{2}\epsilon\,
tuu_x,
\end{eqnarray}
which coincides with the results in \cite{ah}.

Both Eq.(\ref{per-1}) and Eq.(\ref{per-2}) share the same
unperturbed equation (\ref{nonper}), thus formula
(\ref{formula-wave}) provides a direct and simple way to obtain
approximate conservation law based on the known conservation laws
(\ref{un-con}) of Eq.(\ref{nonper}). Moreover, the new formulae
produce more approximate conservation laws than the results stated
in the previous literatures.
%
%This example demonstrates that Theorem \ref{th-3} can be used to
%discriminate the approximate nonlinear self-adjointness of perturbed
%PDEs.

\section{Conclusion and discussion}
We propose a succinct approach to discriminate approximate nonlinear
self-adjointness of perturbed PDEs by means of nonlinear
self-adjointness of the corresponding unperturbed PDEs.
Consequently, an approximate conservation law formula is given based
on the known conservation laws of the unperturbed PDEs, especially
effective for the system containing two or more perturbed PDEs and
sharing the same unperturbed PDEs. The results are applied to
several linear and nonlinear perturbed wave equations and new
approximate conservation laws are obtained.

In addition, it had been shown that Eq.(\ref{short}) is connected
with its adjoint equation by the differential substitution
$v=u_t-\frac{1}{2}u^2u_x$ \cite{ib-2011}, thus there exist some PDEs
which are not approximately nonlinearly self-adjoint in the sense of
Definition \ref{def-1} but with a general definition of
(approximate) nonlinear self-adjointness such as differential
substitution, thus it may consider the extended substitutions such
as differential substitution containing first-order or higher-order
derivatives. We will report these results in the forthcoming papers.


\begin{thebibliography}{99}
\bibitem{ab-1991}M. J. Ablowitz and P. A. Clarkson, \emph{Solitons, Nonlinear Evolution
Equations and Inverse Scattering}, London Mathematical Society
Lecture Note Series 149, Cambridge University Press, Cambridge,
1991.
\bibitem{ma-1988}A.V. Mikhailov, A.B. Shabat and R.I. Yamilov,
Extension of  the  module of  invertible  transformations and
clssification of integrable system,  Comm. in  Math. Phys. 115
(1988) 1-19.
\bibitem{blu1}G.W. Bluman, A.F. Cheviakov and S.C. Anco, \emph{Applications of Symmetry Methods
to Partial Differential Equations}, Springer-Verlag, New York
(2010).
\bibitem{Olver} P.J. Olver, A\emph{pplications of Lie Groups to Differential Equations},
Springer-Verlag, New York (1993).
\bibitem{ba-1996} V.A. Baikov, R.K.
Gazizov, N.H. Ibragimov, in: N.H. Ibragimov (Ed.), \emph{Handbook of
Lie Group Analysis of Differential Equations}, vol. 3, CRC Press,
Boca Raton, FL, 1996.
\bibitem{noether}E. Noether, Invariante
Variationsprobleme, Nachr. K\"{o}nig. Gissell. Wissen.,
G\"{o}ttingen, Math.-Phys. Kl. 2 (1918) 235-257, English transl.:
Transport Theory Statist. Phys. 1 (1971) 186-207.
\bibitem{kara-2006}A.H. Kara, F.M. Mahomed, Noether-type symmetries and conservation laws via partial Lagrangians, Nonlinear
Dyn. 5 (2006) 367-383.
\bibitem{blu11}S.C. Anco and G.W. Bluman,
Direct construction method for conservation laws of partial
differential equations, part I: examples of conservation law
classifications, Eur. J. Appl. Math. 13 (2002) 545-566.
\bibitem{blu21} S.C.  Anco and G.W. Bluman,  Direct construction
method for conservation laws of partial differential equations, part
II: general treatment, Eur. J. Appl. Math. 13 (2002) 567-585.
\bibitem{ba-1931} H. Bateman, On Dissipative Systems and Related Variational
Principles, Phys. Rev. 38 (1931) 815-819.

\bibitem{ca-1988} G. Caviglia, Composite variational principles and the
determination of conservation laws, J. Math. Phys. 29 (1988)
812-816.

\bibitem{nh-2006} N.H. Ibragimov, Integrating factors, adjoint equations and Lagrangians,
J. Math. Anal. Appl. 318 (2006) 742-757.
\bibitem{nh-2007}N.H. Ibragimov, A new conservation theorem, J. Math. Anal. Appl. 333 (2007) 311-328.
\bibitem{ib-2007} N.H. Ibragimov, Quasi self-adjoint differential equations, Arch. ALGA
4 (2007) 55-60.
\bibitem{gan-2011}M.L. Gandarias, Weak self-adjoint
differential equations, J. Phys. A. 44 (2011) 262001 (6pp).

\bibitem{ib-2011} N.H. Ibragimov, Nonlinear self-adjointness in constructing conservation laws, Archives of
ALGA,1-99, Volume 7/8, 2010-2011.
\bibitem{nhi}N.H. Ibragimov,
Nonlinear self-adjointness and conservation laws, J. Phys. A: Math.
Theor. 44 (2011) 432002  (8pp).
\bibitem{tt-2012} M. Torrisi and R. Tracin$\grave{\text{a}}$,
Quasi self-adjointness of a class of third order nonlinear
dispersive equations, Nonlin. Anal. RWA 14 (2013) 1496-1502.
\bibitem{gan-2012} M.L. Gandarias and M. Bruz$\acute{\text{o}}$n, Some
conservation laws for a forced KdV equation, Nonlin. Anal. RWA 13
(2012) 2692-2700.
\bibitem{Fs-2012} I.L. Freire  and J.C.S. Sampaio,
Nonlinear self-adjointness of a generalized fifth-order KdV
equation, J. Phys. A: Math. Theor. 45 (2012)  032001 (7pp).

%\bibitem{nh} N. H. Ibragimov, Handbook of Lie Group Analysis of
%Differential Equations, CRC Press, Boca Raton, vol. 1, 1994; vol. 2,
%1995; vol. 3, 1996.
 \bibitem{nhi-2011}N.H. Ibragimov, M.
Torrisi and R. Tracin$\grave{\text{a}}$, Self-adjointness and
conservation laws of a generalized Burgers equation, J. Phys. A:
Math. Theor. 44 (2011) 145201 (5pp).
\bibitem{fre-2013} I.L. Freire,
New classes of nonlinearly self-adjoint evolution equations of
third- and fifth-order, Commun. Nonlin. Sci. Num. Simul. 18 (2013)
493-499.
\bibitem{ba-1991} V.A. Baikov, R.K.
Gazizov and  N.H. Ibragimov, Perturbation methods in group analysis,
J. Soviet. Math.  55 (1991) 1450-1490.
\bibitem{fs-1989}
W.I. Fushchich and W.H. Shtelen, On approximate symmetry and
approximate solutions of the non-linear wave equation with a small
parameter, J. Phys. A 22 (18) (1989) L887-L890.

\bibitem{pak} M. Pakdemirli, M. Yurusoy and T.
Dolapc, Comparison of Approximate Symmetry Methods for Differential
Equations, Acta Appl. Math. 80 (2004) 243-271.
\bibitem{ron} R. Wiltshire, Two approaches
to the calculation of approximate symmetry exemplilfied using a
system of advection-diffusion equations, J. Comput. Appl. Math.
197(2)(2006) 287-301.
\bibitem{qu1}F.M. Mahomed , C.Z. Qu , Approximate
conditional symmetries for partial differential equations, J. Phys.
A: Math. Gen. 33 (2000) 343-356.
\bibitem{qu2} A.F. Kara, F.M.
Mahomed, C.Z. Qu, Approximate potential symmetries for partial
differential equations, J. Phys. A: Math. Gen. 33 (2000) 6601-6613.
\bibitem{eu-1994}M. Euler, N. Euler,  A. K$\ddot{\mbox{o}}$hler, On the construction of
approximate solutions for a multi-dimensional nonlinear heat
equation, J. Phys. A: Math. Gen. 27 (1994) 2083-2092.
\bibitem{ah} A.G. Johnpillai and A.H.
Kara, Variational Formulation of Approximate Symmetries and
Conservation Laws, Int. J. Theor. Phys. 40 (2001) 1501-1509.
\bibitem{ah-1999}A.H. Kara, F.M. Mahomedl and
G. \"{U}nal, Approximate symmetries and conservation laws with
applications, Int. J. Theoret. Phys. 38 (1999) 2389-2399.

\bibitem{ag-2006}A.G. Johnpillai, A.H. Kara and F.M. Mahomed, A basis of approximate
conservation laws for PDEs with a small parameter, Int. J. Nonlinear
Mech. 41 (2006) 830-837.
\bibitem{ah-2009} A.G. Johnpillai, A.H.
Kara and F.M. Mahomed, Approximate Noether-type symmetries and
conservation laws via partial Lagrangians for PDEs with a small
parameter, J. Comput. Appl. Math. 223 (2009) 508-518.
\bibitem{zhang-2013} Z.Y. Zhang, Approximate nonlinear self-adjointness and approximate conservation
laws.  J. Phys. A: Math. Theor. 46 (2013)  155203
(13pp).
\bibitem{kara-2002} T. Feroze and A.H. Kara, Group theoretic
methods for approximate invariants and Lagrangians for some classes
of $y''+\epsilon F(t)y'+y=f(y,y)$, Int. J. Nonlin. Mech. 37 (2002)
275-280.
\bibitem{sch-2004}T. Sch$\ddot{\mbox{a}}$fer and C. Wayne,
Propagation of ultra-short optical pulses in cubic non- linear
media, Physica D, 196 (2004) 90-105.


 \bibitem{char} I.A. Charnyi,  \emph{Unstable Motion  of a Real Fluid  in Pipes},  Nedra,  Moscow
1975 (in Russian).
 \bibitem{zhang1}Z.Y. Zhang and Y.F. Chen, A
comparative study of approximate symmetry and approximate homotopy
symmetry to a class of perturbed nonlinear wave equation, Nonlinear
Anal. 74 (2011) 4300-4318.




%\bibitem{wf-1981} W.F. Ames,  E. Adams and R.J. Lohner,  Group properties  of $u_{tt} =  [f(u)u_x]_x$, Int. J.
%Non-Linear  Mech. 16 (1981) 439-447.
\end{thebibliography}
\end{document}